\documentclass[aps,pra,reprint,notitlepage,eqsecnum,floatfix,showkeys,nofootinbib,onecolumn]{revtex4-2}

\usepackage{amsmath}
\usepackage{amsbsy}
\usepackage{amsfonts}
\usepackage{xcolor}
\usepackage{graphicx}
\usepackage{mathalpha}
\usepackage{bm}
\usepackage{hyperref}
\usepackage{multirow}
\usepackage{linearb}
\usepackage{relsize}
\usepackage{fancyvrb}
\usepackage{calrsfs}
\DeclareMathAlphabet\mathbfcal{OMS}{cmsy}{b}{n}

\newcommand{\bq}{\begin{eqnarray}}
\newcommand{\eq}{\end{eqnarray}}
\newcommand{\bqn}{\begin{eqnarray*}}
\newcommand{\eqn}{\end{eqnarray*}}
\newcommand{\bqs}{\begin{subequations}}
\newcommand{\eqs}{\end{subequations}}
\newcommand{\bw}{\begin{widetext}}
\newcommand{\ew}{\end{widetext}}

\newcommand{\aaa}{{\boldsymbol a}}
\newcommand{\bbb}{{\boldsymbol b}}

\newcommand{\qq}{{\boldsymbol q}}

\newcommand{\pp}{{\boldsymbol p}}

\newcommand{\bbss}[1]{\boldsymbol{#1}}
\newcommand{\cals}{{\cal S}}

\newcommand{\calp}{\mathbfcal{P}}
\newcommand{\calq}{\mathbfcal{Q}}
\newcommand{\calqq}{{\cal Q}}
\newcommand{\calpp}{{\cal P}}

\newcommand{\cale}{{\cal E}}

\newcommand{\calo}{{\cal O}}

\newcommand{\blue}[1]{{#1}}
\newcommand{\orange}[1]{{#1}}

\begin{document}
\title{Brownian Bridge for Coherent State Path Integral Monte Carlo}

\author{Riccardo Fantoni}
\email{riccardo.fantoni@scuola.istruzione.it}
\affiliation{Universit\`a di Trieste, Dipartimento di Fisica, strada
  Costiera 11, 34151 Grignano (Trieste), Italy}

\date{\today}

\begin{abstract}
We propose a new Brownian bridge construction for our newly devised Coherent States 
Path Integral Monte Carlo algorithm. We apply it to the numerically exact calculation
of the thermodynamic properties of the Helium fluid on a plane at low non zero 
temperature. We find very good agreement with the conventional plane waves path 
integral Monte Carlo results.
\end{abstract}

\keywords{Quantum Many Body; Coherent States; Path Integral; Quantum Monte Carlo; Brownian Bridge; Helium; Thermodynamics}

\maketitle
\section{Introduction}
\label{sec:intro}

We recently \cite{Fantoni25f} constructed a new Path Integral Monte Carlo (PIMC) 
algorithm using Coherent States (CS), \cite{KlauderJCP1963a,KlauderJCP1963b,KlauderJCP1963c,KlauderJCP1963d,KlauderJCP1963e,Klauder-CS} in place of the more usual Plane Waves (PW), as the complete 
set over which to expand the thermal density matrix. We called this 
algorithm with the acronym CSPIMC. We proposed our new algorithm as
an alternative to the more common Plane Waves PIMC of Ref. \cite{Ceperley1995}
to which we will refer with the acronym PWPIMC. 

Although the mathematics of Ref. \cite{Fantoni25f} where we proposed the 
new algorithm is completely correct, the numerics in that paper was not 
complete. In particular it was lacking a justification of the Brownian bridges 
\cite{Kalos-Whitlock,Ceperley1995} used in the CSPIMC scheme to sample the 
permutations sum necessary to calculate the properties of identical particles 
(being them bosons or fermions). 

In the present work we will fill that gap proposing a correct realization of 
the Brownian bridge move within the CSPIMC algorithm. And we will justify our 
new proposal.

The CSPIMC numerical results that we obtain for Helium on a plane compare
well with the results from the conventional PWPIMC. Within CSPIMC the 
continuum limit for the imaginary time must be taken simultaneously to the 
limit of an infinite elastic constant of the harmonic oscillator of unit 
mass subtending the coherent states.

\section{Physics model}
\label{sec:intro}

In Ref. \cite{Fantoni25f} we performed several Path Integral Monte Carlo 
(PIMC) simulations for a two dimensional, $d=2$, $^4$He liquid 
\cite{Gordillo1998} with either Boltzmann or Bose statistics. 

The liquid has a surface number density $n=N/\Omega$ where $N$ is the number of 
Helium atoms in an area $\Omega$ of a flat surface, at an inverse temperature 
$\beta=1/k_BT$ with $k_B$ Boltzmann constant. The Hamiltonian of the fluid is
$\hat{H}=\hat{T}+\upsilon\hat{V}=\hat{P}^2/2m+\upsilon V(Q)$ with
$\vec{Q}=(\qq_1,\qq_2,\ldots,\qq_N)$ the particles positions and 
$\vec{P}=(\hat{\pp}_1,\hat{\pp}_2,\ldots,\hat{\pp}_N)=
-i(\bbss{\nabla}_{\qq_1},\bbss{\nabla}_{\qq_2},\ldots,\bbss{\nabla}_{\qq_N})$
the particles momenta. The potential energy $V(Q)$ enters the Hamiltonian 
with a coupling constant $\upsilon$.

The computer experiments \cite{Fantoni25f} were done for for $N=16$ $^4$He atoms 
($m=0.0830594\approx 1/12~\mbox{\AA}^{-2}\mbox{K}^{-1}$) in two dimensions $d=2$, 
in a square periodic cell of area $\Omega=L^2$, interacting with a Lennard-Jones 
pair potential with parameters \cite{McMillan1964} $\sigma=2.556~\mbox{\AA}$, 
$\varepsilon=10.22~\mbox{K}$ and a cutoff distance $r_{\rm cut}=2.5~\mbox{\AA}$ (so that 
$v(r)=0$ for $r>r_{\rm cut}\sigma$ without long range corrections), at a number density 
$n=N/L^2=0.05~\mbox{\AA}^{-2}$ and various temperatures $T$ or timeslices 
number $M$, either with Boltzmann and Bose statistics. As we can see from the 
phase diagram of Ref. \cite{Gordillo1998}, at this density the fluid undergoes 
a phase transition from a fluid phase at high temperature to a superfluid phase 
at low temperatures.

In that work \cite{Fantoni25f} we compared two different PIMC simulations: 
the conventional Plane Waves PIMC ({\bf PW}PIMC) \cite{Ceperley1995} and our 
newly devised Coherent States PIMC ({\bf CS}PIMC) algorithm. The two PIMC 
differ for the expression of the hot kinetic density matrix at an imaginary 
timestep $\tau=\beta/M$ with $M$ a large number of timeslices. The short 
imaginary time density matrix CSPIMC expression is reviewed in the next section.
The PWPIMC requires a multidimensional integral over $dNM$
coordinates whereas our CSPIMC requires $5dMN$ integrations, the usual $dNM$
real particles coordinates, $2dNM$ ghost particles coordinates, and $2dNM$
ghost particles momenta. The coherent states are generated by a harmonic
oscillator (h.o.) with mass $m_{h.o.}$ and elastic constant $k=m_{h.o.}\omega^2$.
We introduced the parameter $\xi\equiv m_{h.o.}\omega/2$ and the adimensional 
one $\varphi\equiv\xi\tau/m=(\sigma_{p.w.}/\sigma_{c.s.})^2/2$ where
$\sigma_{p.w.}\equiv\sqrt{2\lambda\tau}$ is the standard deviation in the 
plane wave scheme, with $\lambda=1/2m$, and 
$\sigma_{c.s.}\equiv\sqrt{1/m_{h.o.}\omega}$ is the standard deviation in the
coherent state scheme. As illustrated in Ref. \cite{Fantoni25f} in order
to compare PWPIMC and CSPIMC it is necessary to keep $\varphi$ fixed.

In our quantum simulations we have to take care of two limiting 
procedures: The continuum limit where the ultraviolet cutoff $\tau\to 0$ or 
$M\to\infty$ at fixed absolute temperature $T$ and the thermodynamic limit 
where $N\to\infty$ and $\Omega\to\infty$ at fixed density $n$. In particular 
in Ref. \cite{Fantoni25f} we explicitly worried about the continuum limit and 
we proved that in the CSPIMC scheme it is necessary to keep $\varphi$
constant as $\tau\to 0$ so that $\xi\to\infty$. To mimic the 
thermodynamic limit we simply used a periodic square cell which permeates 
the whole infinite space. This procedure gives rise to spurious finite size 
effects unavoidable on a computer.

We worked in units where $\hbar=k_B=1$ so that the
imaginary time has dimensions of temperature.

\begin{widetext}
\section{The short time density matrix for CSPIMC}
\label{sec:stgf}

The short imaginary time, $\tau$, Green function for CSPIMC can be written as 
follows \cite{Fantoni25f}:
\bq \label{eq:K}
\rho(Q,Q';\tau)\approx
e^{-\tau\blue{\upsilon}V(Q)}\frac{1}{N!}\sum_P
\prod_{\alpha=1}^N\zeta_\alpha\Bigl(\qq_{P\alpha}|\qq'_\alpha;\tau,m,\xi\Bigr),
\eq
with $P$ a permutation of the $N$ {\bf real} identical particles and
\bq
\zeta_\alpha\Bigl(\qq|\qq';\tau,m,\xi\Bigr)
\approx\int\frac{d\calq_\alpha\,d\calp_\alpha}{(2\pi)^d}
\frac{d\calq'_\alpha\,d\calp'_\alpha}{(2\pi)^d}
\psi_\alpha(\qq_\alpha,\calq_\alpha,\calp_\alpha)
{\psi_\alpha}^*(\qq'_\alpha,\calq'_\alpha,\calp'_\alpha)G_\alpha\exp
\left[-\tau\orange{(\calp_\alpha^2+{\calp'}_\alpha^2)/4m}\right],
\eq
where $\psi_\alpha(\qq_\alpha,\calq_\alpha,\calp_\alpha)$ is the CS wave 
function with ghost canonical variables $\calq_\alpha$ and $\calp_\alpha$, the 
{\bf ghost} of the real particle $\alpha$, namely
\bq \label{eq:csq}
\psi_\alpha(\qq_\alpha,\calq_\alpha,\calp_\alpha)
&\equiv&\langle\qq_\alpha|\calq_\alpha,\calp_\alpha\rangle
=\left(\frac{2\xi}{\pi}\right)^{d/4}\exp\left[-\xi
\left(\qq_\alpha-\calq_\alpha\right)^2+i\calp_\alpha\cdot
\left(\qq_\alpha-\calq_\alpha\right)\right],\\
G_\alpha&=&\exp\left[{-\frac{1}{2}(|\aaa|^2+|\bbb|^2)+
\aaa^*\cdot \bbb+\frac{i}{2}(\calq_\alpha\cdot\calp_\alpha-
\calq'_\alpha\cdot\calp'_\alpha)}\right],\\ \label{eq:i}
\aaa&\equiv&\frac{1}{\sqrt{4\xi}}(2\xi\calq_\alpha+i\calp_\alpha),\\
\bbb&\equiv&\frac{1}{\sqrt{4\xi}}(2\xi\calq'_\alpha+i\calp'_\alpha),
\eq
where $G_\alpha$ is the scalar product of the two ghosts $\calq$ and $\calq'$. 
\end{widetext}

We can then define a reduced high temperature density matrix
\bq
\tilde{\rho}(s,s';\tau)
&\propto&e^{-\tau\blue{\upsilon}V(Q)}
\prod_{\alpha=1}^N\psi_\alpha(\qq_\alpha,\calq_\alpha,\calp_\alpha)
{\psi_\alpha}^*(\qq'_\alpha,\calq'_\alpha,\calp'_\alpha)G_\alpha 
e^{-\tau\orange{(\calp_\alpha^2+{\calp'}_\alpha^2)/4m}},
\eq
with $s=(\vec{Q},\vec{\calqq},\vec{\calpp},P)$ the configurations space
for a single link where $\vec{\calqq}=(\calq_1,\calq_2,\ldots,\calq_N)$ and 
$\vec{\calpp}=(\calp_1,\calp_2,\ldots,\calp_N)$.

Discretizing the imaginary time into $M$ timeslices with $\tau=\beta/M$ we can 
use Trotter \cite{Trotter1959} identity to introduce $M$ links 
$\tilde{\rho}(s_k,s_{k+1};\tau)$ for $k=0,1,2,\ldots,M-1$ to reconstruct the 
density matrix at an inverse temperature $\beta$ as the following path integral
\bq \nonumber
\rho(s_0,s_M';\beta)&\propto&
\sum_P\int\tilde{\rho}(s_0,s_1';\tau)\tilde{\rho}(s_1,s_2';\tau)\cdots
\tilde{\rho}(s_{M-1},s_M';\tau)\times\\ \label{eq:dm}
&&\prod_{\alpha=1}^N\prod_{k=1}^{M-1}
d\qq_{\alpha,k}\frac{d\calq_{\alpha,k-1}^{k}d\calp_{\alpha,k-1}^{k}}{(2\pi)^d}
\frac{d\calq_{\alpha,k}^{k}d\calp_{\alpha,k}^{k}}{(2\pi)^d}
\frac{d\calq_{\alpha,M-1}^{M}d\calp_{\alpha,M-1}^{M}}{(2\pi)^d}
\frac{d\calq_{\alpha,M}^{M}d\calp_{\alpha,M}^{M}}{(2\pi)^d}
\eq
where $\qq_{\alpha,k}$ are the coordinates of the real particle $\alpha$ at
timeslice $k$, $\calq_{\alpha,k}^l$ and $\calp_{\alpha,k}^l$ are respectively 
the canonical coordinates and momenta of the ghost of particle $\alpha$ at the 
timeslice $k$ in the link $l-1\to l$. In our notation
$s_k=(\vec{Q}_k,\vec{\calqq}_k^{k+1},\vec{\calpp}_k^{k+1},P)$ and
$s_k'=(\vec{Q}_k,\vec{\calqq}_k^k,\vec{\calpp}_k^k,P)$ with 
$\vec{Q}_k=(\qq_{1,k},\qq_{2,k},\ldots,\qq_{N,k})$,
$\vec{\calqq}_k^l=(\calq_{1,k}^l,\calq_{2,k}^l,\ldots,\calq_{N,k}^l)$, and
$\vec{\calpp}_k^l=(\calp_{1,k}^l,\calp_{2,k}^l,\ldots,\calp_{N,k}^l)$.

The Monte Carlo used is the standard Metropolis algorithm 
\cite{Metropolis,Kalos-Whitlock}. In our CSPIMC simulations we 
have two kinds of moves: a singleslice {\sl displace move} and
a multislice {\sl Brownian bridge}. 

In the displace move we choose 
as transition move a uniform displacement of each of the $dMN$ 
real path coordinate 
$\qq_{\alpha,k}\to\qq_{\alpha,k}+(1/2-\eta)\mathbf{\Delta}$ for 
$\alpha=1,\ldots,N$ and $k=1,\ldots,M$, where $\eta$ is a uniform 
pseudorandom number in $[0,1)$ and $\mathbf{\Delta}$ a fixed 
$d$-dimensional vector whose magnitude is chosen so to have
acceptance ratios close to $1/2$. And of each of the $4dMN$ ghost path 
canonical variables
$\calq_{\alpha,k-1}^k\to\calq_{\alpha,k-1}^k+(1/2-\eta)\mathbf{\Delta}$, 
$\calp_{\alpha,k-1}^k\to\calp_{\alpha,k-1}^k+(1/2-\eta)\mathbf{\Delta}$, and
$\calq_{\alpha,k}^k\to\calq_{\alpha,k}^k+(1/2-\eta)\mathbf{\Delta}$, 
$\calp_{\alpha,k}^k\to\calp_{\alpha,k}^k+(1/2-\eta)\mathbf{\Delta}$. 
So that the transition probability density is just a constant and drops 
out of the acceptance probability.

The sum over permutations in Eq. (\ref{eq:dm}) can be applied just to the real 
particles coordinates in the first (or last) timeslice. In order to sample the 
permutation sum we need to construct two Brownian bridges 
\cite{Kalos-Whitlock,Ceperley1995} between the initial positions 
taken from the paths of two particles $\alpha$ and $\gamma$ at the same 
timeslice and the final positions taken from the same two paths at a 
subsequent timeslice but exchaged, $\gamma$ and $\alpha$, so to connect 
particle $\alpha$ at the initial timeslice to particle $\gamma$ at the 
final timeslice with one bridge and particle $\gamma$ at the initial 
timeslice with particle $\alpha$ at the final timeslice. We 
accept or reject the move according to Metropolis algorithm
\cite{Metropolis,Kalos-Whitlock}. If the move
is accepted one creates an exchange of two particles. Since any permutation
of $N$ particles can be obtained by composing a finite number of  
particles exchanges this is sufficient to simulate Bose-Einstein statistics
in an exact numerical way. The Brownian bridge move is described in the next 
section.

In our simulations we measured the total potential energy 
${\cal E}_P=\upsilon\langle V\rangle$ and the total kinetic energy
\bq \label{eq:ke}
{\cal E}_K=\left\{\begin{array}{ll} \displaystyle
\frac{dN}{2\tau} - \frac{\langle(\vec{Q}_k-\vec{Q}_{k-1})^2\rangle}
{4\lambda\tau^2} &\mbox{PWPIMC}\\ \displaystyle
\frac{dN}{2\tau}\frac{\varphi}{1+\varphi} - \mu\frac{\langle\vec{\calpp}_k^2+
\vec{\calpp}_k^{\prime 2}\rangle}{4m} &\mbox{CSPIMC}
\end{array}\right.
\eq
where, given an observable $\calo$, we denote with 
$\langle\calo\rangle=\mbox{tr}(\rho\calo)/\mbox{tr}(\rho)$ 
the thermal average with `tr' the trace operation. These 
expressions for $\cale_K$ and $\cale_P$ corresponds to the 
{\sl thermodynamic} estimators of Ref. \cite{Ceperley1995}. 
$\cale=\cale_K+\cale_P$ is the total internal energy. The parameter 
$\mu$, in the CSPIMC case, is necessary in order to find 
agreement with the kinetic energy of the PWPIMC, as explained in Ref. 
\cite{Fantoni25f}. The kinetic energy from Eq. (\ref{eq:ke}) is the
small difference between two large quantities, infinite in the continuum 
$\tau\to 0$ limit. Note that in the $\xi\to\infty$ limit
$\varphi/(1+\varphi)\to 1$ in agreement with the PWPIMC case.

\section{The bridge move}
\label{sec:bridge}

In order to take into account the particles permutations it is necessary
to construct two Brownian bridges between two different
\footnote{A bridge between the same particle can still be used to sample the
density matrix of distinguishable particles as can be done with the displacement
move.}
randomly chosen particles in two randomly chosen many body beads to generate 
an exchange between the two particles. With one bridge we connect a real 
particle $1$ and its ghosts on timeslice $i$ to particle $2$ and its ghosts on 
timeslice $j$ and with the other we connect particle $2$ and its ghosts on timeslice
$i$ to particle $1$ and its ghosts on timeslice $j$ with $i<j$. 
This will produce an {\sl exchange} of particles 1 and 2 and their respective 
ghosts.

The Brownian bridge between particle $1$ at $\qq_{1,i}$ and particle $2$ at 
$\qq_{2,j}$ is built through the following multislice transition move  
\cite{Ceperley1995},
\bq \label{eq:bmove1}
\qq_{{\rm new},i}&=&\qq_{1,i}\\ \label{eq:bmove2}
\qq_{{\rm new},k}&=&\qq_{{\rm new},k-1}+
\frac{\qq_{2,j}-\qq_{{\rm new},k-1}}{j-k+1}+
\xi~~~~~~k=i+1,\ldots,j
\eq
where $\xi$ is a random number with a Gaussian probability distribution
\footnote{This can be generated with the Box-Muller algorithm 
\cite{Kalos-Whitlock} for example.}
with zero mean and variance $\sigma^2(j-k)/(j-k+1)$ where 
$\sigma^2=2\lambda\tau$ is the diagonal free particle variance. The same 
move (\ref{eq:bmove1})-(\ref{eq:bmove2}) is also carried out on the ghosts
coordinates $\calq_{\alpha,k}^k$ and $\calq_{\alpha,k}^{k+1}$ and we leave 
untouched the momenta of the two ghosts. 

The Metropolis (rejection) method \cite{Metropolis} can sample any probability 
distribution provided that the transition rule satisfies detailed balance and 
ergodicity. The Metropolis algorithm is a particular way of ensuring that
the transition rule satisfies detailed balance. It does this by
splitting the {\sl transition probability} into an ``a priori'' {\sl sampling
distribution} $T(S_{\rm old}\to S_{\rm new})$ (which is a probability 
distribution that we already know) and an {\sl acceptance probability} 
$A(S_{\rm old}\to S_{\rm new})$.
\bq
P(S_{\rm old}\to S_{\rm new})=T(S_{\rm old}\to S_{\rm new})
A(S_{\rm old}\to S_{\rm new}),
\eq
where we denote with $S\equiv (\{s_k\},\{s_k'\})$, $S_{\rm old}$ is the
state configuration before the Markov move and $S_{\rm new}$ the proposed
configuration state after the move produced by the transition with probability
$P$. 

In the generalized Metropolis procedure \cite{Kalos-Whitlock},
trial moves are accepted according to:
\bq \label{eq:acceptance}
A(S_{\rm old}\to S_{\rm new})=\min[1,q(S_{\rm old}\to S_{\rm new})],
\eq
where
\bq
q(S_{\rm old}\to S_{\rm new})=\frac{\pi(S_{\rm new})T(S_{\rm new}\to S_{\rm old})}
{\pi(S_{\rm old})T(S_{\rm old}\to S_{\rm new})}.
\eq
where $\pi\propto e^{-\cals}\propto\rho(s_0,s_1';\tau)\rho(s_1,s_2';\tau)\cdots
\rho(s_{M-1},s_M';\tau)$ is the action probability distribution in 
the $s=(\{s_k\})$ configurations space. The transition probability 
corresponding to the move of Eqs. (\ref{eq:bmove1})-(\ref{eq:bmove2}) is 
then given by
\bq \nonumber
\frac{T(S_{\rm new}\to S_{\rm old})}{T(S_{\rm old}\to S_{\rm new})}
\propto&\exp&\left\{\frac{1}{4\lambda\tau}\sum_{k=i+1}^{j}\left[
(\qq_{{\rm new},k}-\qq_{{\rm new},k-1})^2+
(\calq_{{\rm new},k}^k-\calq_{{\rm new},k-1}^k)^2+
(\calq_{{\rm new},k}^{k+1}-\calq_{{\rm new},k-1}^{k+1})^2\right.\right.\\
&&\left.\left.-(\qq_{{\rm old},k}-\qq_{{\rm old},k-1})^2-
(\calq_{{\rm old},k}^k-\calq_{{\rm old},k-1}^k)^2-
(\calq_{{\rm old},k}^{k+1}-\calq_{{\rm old},k-1}^{k+1})^2\right]
\phantom{\frac{1}{2}}\right\}.
\eq

In order to produce an exchange of two particles $1$ and $2$ one needs a 
combination of two bridge transitions as described above. Together with 
the real particles we propose also the exchange of the relative 
ghosts. 
Any permutation can be reached through a two particles exchange so the 
bridge transition move allows to sample the sum in Eq. (\ref{eq:dm}).

In order to increase the acceptance ratios of this bridge and swap moves 
it is necessary to diminish $\xi$ and the number $j-i$ of its timeslices. 
Of course our kinetic energy estimator is left unchanged by this bridge 
move since the ghosts momenta are not moved at all. But the potential 
energy obtained with only the bridge move converges to the same value 
obtained with the displace and the bridge moves together, as it should. 
Moreover the potential energy becomes more negative as $\varphi$ increases
keeping all other conditions constant.

\section{Numerical results}
\label{sec:results}

We here compare the conventional PWPIMC with our CSPIMC algorithm on specific simulations.

Our PWPIMC simulations (see Tables \ref{tab:pw-boltzmann-tau}, \ref{tab:pw-bose-tau} and 
\ref{tab:pw-boltzmann}, \ref{tab:pw-bose}) confirm that at high temperature (classical 
regime) the nature of the statistics is not important. Whereas 
it becomes important at low temperatures (quantum regime). The zero temperature (ground
state) limit can only be reached through an extrapolation of the PIMC results.
In Tables \ref{tab:pw-boltzmann-tau} and \ref{tab:pw-bose-tau} we use PWPIMC at constant
$\tau=0.025~\mbox{K}^{-1}$ (same as Ref. \cite{Gordillo1998}) and various temperatures.
In Tables \ref{tab:pw-boltzmann} and \ref{tab:cs-boltzmann} we compare the PWPIMC with the 
CSPIMC for Boltzmann statistics at fixed $M=250$ and various temperatures.
In Tables \ref{tab:pw-bose} and \ref{tab:cs-bose} we do the same for Bose statistics.
We found favorable match between PWPIMC and CSPIMC results for 
$\varphi=0.7\approx 2^{-1/2}$. 
\footnote{Note that in Ref. \cite{Fantoni25f} we fixed $\varphi=\pi/2$ but the
comparison between CSPIMC and PWPIMC results was not very good. Moreover here we
changed the acceptance probabilities of both the displace and the bridge moves.}
Each run has no less than $2\times 10^7$ MC steps where one step is made of a displace 
move of all the timeslices of a single particle path and their associated ghost paths 
and a bridge move.

\begin{table}[hbt]
\caption{Results from {\bf PW}PIMC for $N=16$ $^4$He atoms ($m=0.0830594$) in two 
dimensions, with a Lennard-Jones pair potential with parameters \cite{McMillan1964} 
$\sigma=2.556, \varepsilon=10.22$ and a cutoff distance $r_{\rm cut}=2.5$ (so that 
$v(r)=0$ for $r>r_{\rm cut}\sigma$ without long range corrections), at a density 
$N/L^2=0.05$ and various temperatures $T$ with $\tau=0.025$, with {\bf Boltzmann statistics}. 
In the Table ${\cal E}_K$ and ${\cal E}_P$ 
are the total kinetic and potential energies respectively.
${\cal E}={\cal E}_K+{\cal E}_P$ is the total internal energy.} 
\label{tab:pw-boltzmann-tau}
\begin{center}
\begin{tabular}{||c||c||c||c||}
\hline
$T~(\mbox{K})$ & ${\cal E}_K~(\mbox{K})$ & $-{\cal E}_P~(\mbox{K})$ & $-{\cal E}~(\mbox{K})$\\
\hline
\hline
1.0 & $66.6(2)$ & $94.4(1)$   & $27.8$ \\
0.5 & $60.8(2)$ & $92.9(1)$   & $32.1$ \\
0.2 & $59.0(2)$ & $93.00(8)$  & $34.0$ \\
0.1 & $58.2(2)$ & $93.23(8)$  & $35.0$\\
\hline
\end{tabular}
\end{center}
\end{table}
\begin{table}[hbt]
\caption{Results from {\bf PW}PIMC for $N=16$ $^4$He atoms ($m=0.0830594$) in two 
dimensions, with a Lennard-Jones pair potential with parameters \cite{McMillan1964} 
$\sigma=2.556, \varepsilon=10.22$ and a cutoff distance $r_{\rm cut}=2.5$ (so that 
$v(r)=0$ for $r>r_{\rm cut}\sigma$ without long range corrections), at a density 
$N/L^2=0.05$ and various temperatures $T$ with $\tau=0.025$, with {\bf Bose statistics}. 
In the Table ${\cal E}_K$ and ${\cal E}_P$ 
are the total kinetic and potential energies respectively.
${\cal E}={\cal E}_K+{\cal E}_P$ is the total internal energy.} 
\label{tab:pw-bose-tau}
\begin{center}
\begin{tabular}{||c||c||c||c||}
\hline
$T~(\mbox{K})$ & ${\cal E}_K~(\mbox{K})$ & $-{\cal E}_P~(\mbox{K})$ & $-{\cal E}~(\mbox{K})$\\
\hline
\hline
1.0 & $57.1(1)$ & $94.8(3)$   & $37.7$ \\
0.5 & $55.0(5)$ & $93.4(2)$   & $38.4$ \\
0.2 & $55.8(1)$ & $93.55(5)$  & $37.7$ \\
0.1 & $55.9(1)$ & $93.46(5)$  & $37.6$ \\
\hline
\end{tabular}
\end{center}
\end{table}
\begin{table}[hbt]
\caption{Results from {\bf PW}PIMC for $N=16$ $^4$He atoms ($m=0.0830594$) in two 
dimensions, with a Lennard-Jones pair potential with parameters \cite{McMillan1964} 
$\sigma=2.556, \varepsilon=10.22$ and a cutoff distance $r_{\rm cut}=2.5$ (so that 
$v(r)=0$ for $r>r_{\rm cut}\sigma$ without long range corrections), at a density 
$N/L^2=0.05$ and various temperatures $T$ with $M=250$, with {\bf Boltzmann statistics}. 
The runs are $2\times 10^7$ steps long. In the Table ${\cal E}_K$ and ${\cal E}_P$ 
are the total kinetic and potential energies respectively.
${\cal E}={\cal E}_K+{\cal E}_P$ is the total internal energy (note that unlike Ref.
\cite{Gordillo1998} we fix $M$ and not $\tau$).} 
\label{tab:pw-boltzmann}
\begin{center}
\begin{tabular}{||c||c||c||c||}
\hline
$T~(\mbox{K})$ & ${\cal E}_K~(\mbox{K})$ & $-{\cal E}_P~(\mbox{K})$ & $-{\cal E}~(\mbox{K})$\\
\hline
\hline
1.0 & $82.3(6)$ & $83.6(1)$   & $1.3(6)$ \\
0.5 & $74.6(4)$ & $84.8(1)$   & $10.2(4)$\\
0.2 & $62.5(4)$ & $90.5(1)$   & $28.0(4)$\\
0.1 & $49.0(2)$ & $101.1(1)$  & $52.1(2)$\\
\hline
\end{tabular}
\end{center}
\end{table}
\begin{table}[hbt]
\caption{Results from {\bf PW}PIMC for $N=16$ $^4$He atoms ($m=0.0830594$) in two 
dimensions, with a Lennard-Jones pair potential with parameters \cite{McMillan1964} 
$\sigma=2.556, \varepsilon=10.22$ and a cutoff distance $r_{\rm cut}=2.5$ (so that 
$v(r)=0$ for $r>r_{\rm cut}\sigma$ without long range corrections), at a density 
$N/L^2=0.05$ and various temperatures $T$ with $M=250$, with {\bf Bose statistics}. 
The runs are $2\times 10^7$ steps long. In the Table ${\cal E}_K$ and ${\cal E}_P$ 
are the total kinetic and potential energies respectively.
${\cal E}={\cal E}_K+{\cal E}_P$ is the total internal energy (note that unlike Ref.
\cite{Gordillo1998} we fix $M$ and not $\tau$).} 
\label{tab:pw-bose}
\begin{center}
\begin{tabular}{||c||c||c||c||}
\hline
$T~(\mbox{K})$ & ${\cal E}_K~(\mbox{K})$ & $-{\cal E}_P~(\mbox{K})$ & $-{\cal E}~(\mbox{K})$\\
\hline
\hline
1.0 & $77(1)$   & $84.1(4)$  & $7.1(4)$\\
0.5 & $69.7(7)$ & $84.6(4)$  & $14.9(7)$\\
0.2 & $60.0(5)$ & $90.6(1)$  & $30.6(5)$\\
0.1 & $49.3(3)$ & $101.1(2)$ & $51.8(3)$\\
\hline
\end{tabular}
\end{center}
\end{table}
\begin{table}[hbt]
\caption{Results from {\bf CS}PIMC with $\varphi=0.7$ for $N=16$ $^4$He atoms 
($m=0.0830594$) in two 
dimensions, with a Lennard-Jones pair potential with parameters \cite{McMillan1964} 
$\sigma=2.556, \varepsilon=10.22$ and a cutoff distance $r_{\rm cut}=2.5$ (so that 
$v(r)=0$ for $r>r_{\rm cut}\sigma$ without long range corrections), at a density 
$N/L^2=0.05$ and various temperatures $T$ with $M=250$, with {\bf Boltzmann statistics}. 
The runs are $2\times 10^7$ steps long. In the Table ${\cal E}_K$ is taken from 
Table \ref{tab:pw-boltzmann} and used to determine $\mu$. ${\cal E}_P$ 
is the total potential energy.} 
\label{tab:cs-boltzmann}
\begin{center}
\begin{tabular}{||c||c||c||c||c||}
\hline
$T~(\mbox{K})$ & $\langle\vec{\calpp}_k^2+\vec{\calpp}_k^{\prime 2}\rangle/4m~(\mbox{K})$ & ${\cal E}_K~(\mbox{K})$ & $-{\cal E}_P~(\mbox{K})$ & $\mu$\\
\hline
\hline
1.0 & $3550(6)$   & $82.3$ & $84.4(4)$  & $0.44$\\
0.5 & $1608(3)$   & $74.6$ & $85.4(2)$  & $0.47$\\
0.2 & $607(1)$    & $62.5$ & $90.3(2)$  & $0.44$\\
0.1 & $293.3(1)$  & $49.0$ & $99.1(1)$ & $0.40$\\
\hline
\end{tabular}
\end{center}
\end{table}
\begin{table}[hbt]
\caption{Results from {\bf CS}PIMC with $\varphi=0.7$ for $N=16$ $^4$He atoms 
($m=0.0830594$) in two 
dimensions, with a Lennard-Jones pair potential with parameters \cite{McMillan1964} 
$\sigma=2.556, \varepsilon=10.22$ and a cutoff distance $r_{\rm cut}=2.5$ (so that 
$v(r)=0$ for $r>r_{\rm cut}\sigma$ without long range corrections), at a density 
$N/L^2=0.05$ and various temperatures $T$ with $M=250$, with {\bf Bose statistics}. 
The runs are $2\times 10^7$ steps long. In the Table ${\cal E}_K$ is taken from 
Table \ref{tab:pw-bose} and used to determine $\mu$. ${\cal E}_P$ 
is the total potential energy.} 
\label{tab:cs-bose}
\begin{center}
\begin{tabular}{||c||c||c||c||c||}
\hline
$T~(\mbox{K})$ & $\langle\vec{\calpp}_k^2+\vec{\calpp}_k^{\prime 2}\rangle/4m~(\mbox{K})$ & ${\cal E}_K~(\mbox{K})$ & $-{\cal E}_P~(\mbox{K})$ & $\mu$\\
\hline
\hline
1.0 & $3721(5)$     & $77$   & $82.6(2)$   & $0.42$ \\
0.5 & $1611(3)$     & $69.7$ & $86.4(2)$   & $0.48$ \\
0.2 & $606.8(8)$    & $60.0$ & $89.8(1)$   & $0.44$ \\
0.1 & $293.0(1)$    & $49.3$ & $100.42(6)$ & $0.40$ \\
\hline
\end{tabular}
\end{center}
\end{table}

In Tables \ref{tab:Mcs-boltzmann} and \ref{tab:Mcs-bose} we performed simulations 
at fixed $\varphi=0.7$, and fixed temperatures $T=1~(\mbox{K})$ and $T=0.2~(\mbox{K})$ 
respectively, at increasing numbers $M$ of timeslices. 
We see that it is necessary to keep $\varphi$ constant upon taking the continuum 
limit, $\tau\to 0$, in order to find agreement between the results for the CSPIMC 
and the ones for the PWPIMC for both the kinetic and potential energies. 
We find that, in the continuum limit, $\mu$ tends to $\approx 0.4$.

\begin{table}[hbt]
\caption{Results from {\bf CS}PIMC with $\varphi=0.7$ for $N=16$ $^4$He atoms 
($m=0.0830594$) in two 
dimensions, with a Lennard-Jones pair potential with parameters \cite{McMillan1964} 
$\sigma=2.556, \varepsilon=10.22$ and a cutoff distance $r_{\rm cut}=2.5$ (so that 
$v(r)=0$ for $r>r_{\rm cut}\sigma$ without long range corrections), at a density 
$N/L^2=0.05$ and a temperature $T=0.2$ with various values of $M$, with {\bf Boltzmann 
statistics}. In the Table ${\cal E}_K\approx 62.5$ is used to determine $\mu$. 
${\cal E}_P$ is the total potential energy. The PWPIMC at $M=1000$ gives an 
$\cale_P=-83.64(7)$.} 
\label{tab:Mcs-boltzmann}
\begin{center}
\begin{tabular}{||c||c||c||c||}
\hline
$M$ & $\langle\vec{\calpp}_k^2+
\vec{\calpp}_k^{\prime 2}\rangle/4m~(\mbox{K})$ & $\mu$ & $-{\cal E}_P~(\mbox{K})$ \\
\hline
\hline
250  & $607(1)$    & $0.44$  & $90.3(2)$ \\
500  & $1268(1)$   & $0.47$  & $86.7(1)$ \\
750  & $2208(3)$   & $0.42$  & $88.0(1)$ \\
1000 & $4156(4)$   & $0.30$  & $86.0(3)$ \\
\hline
\end{tabular}
\end{center}
\end{table}
\begin{table}[hbt]
\caption{Results from {\bf CS}PIMC with $\varphi=0.7$ for $N=16$ $^4$He atoms 
($m=0.0830594$) in two 
dimensions, with a Lennard-Jones pair potential with parameters \cite{McMillan1964} 
$\sigma=2.556, \varepsilon=10.22$ and a cutoff distance $r_{\rm cut}=2.5$ (so that 
$v(r)=0$ for $r>r_{\rm cut}\sigma$ without long range corrections), at a density 
$N/L^2=0.05$ and a temperature $T=0.2$ with various values of $M$, with {\bf Bose 
statistics}. In the Table ${\cal E}_K\approx 60.0$ is used to determine $\mu$.
${\cal E}_P$ is the total potential energy. The PWPIMC at $M=1000$ gives an 
$\cale_P=-84.0(1)$.} 
\label{tab:Mcs-bose}
\begin{center}
\begin{tabular}{||c||c||c||c||}
\hline
$M$ & $\langle\vec{\calpp}_k^2+
\vec{\calpp}_k^{\prime 2}\rangle/4m~(\mbox{K})$ & $\mu$ & $-{\cal E}_P~(\mbox{K})$ \\
\hline
\hline
250  & $606.8(8)$  & $0.44$  & $89.8(1)$ \\
500  & $1272(1)$   & $0.47$  & $86.2(1)$ \\
750  & $2241(5)$   & $0.41$  & $87.8(2)$ \\
1000 & $4193(3)$   & $0.30$  & $85.2(2)$ \\
\hline
\end{tabular}
\end{center}
\end{table}
%

\section{Conclusions}
\label{sec:conclusions}

In this short work we proposed and tested a possible realization of the Brownian
bridge move in the CSPIMC algorithm presented in our recent work of Ref.
\cite{Fantoni25f}. As discussed in that work in the CSPIMC algorithm  
is necessary to take the continuum limit at fixed $\varphi=\xi\tau/m$
by letting $\tau\to 0$ and $\xi\to\infty$. For non zero $\tau$, we found 
agreement between our CSPIMC and the conventional PWPIMC for 
$\varphi\approx 2^{-1/2}$.
\footnote{This corrects the value $\approx\pi/2$ used in Ref. \cite{Fantoni25f}.}
So that
\bq \nonumber
\mbox{CSPIMC} \overset{\varphi\to 1/\sqrt 2}{\xrightarrow{\hspace{3cm}}}\mbox{PWPIMC},
\eq
where the CSPIMC is the PIMC proposed in Ref. \cite{Fantoni25f} and 
PWPIMC is the PIMC defined in Ref. \cite{Ceperley1995}.

We explicitly proved numerically that the simulation with only the displace move 
coincides with the simulation with only the bridge move between two 
timeslices of the same particle path. In these cases one is simulating particles 
obeying to Boltzmann statistics. For bosons particles we proved that the 
full simulation with both displace and bridge moves coincides with the 
simulation with only the bridge move between two timeslices of two different
particles paths allowing therefore two particles exchanges.
This holds even if in the bridge we do not move the two ghosts momenta.

The necessary relation
\bq \nonumber
\xi=\frac{m}{\sqrt{2}\tau},
\eq 
tells that in the continuum limit $\tau\to 0$ the unit mass, $m_{h.o.}=1$, 
harmonic oscillator underlying the coherent states tends to become 
infinitely stiff with an elastic constant $k=4\xi^2\to\infty$. 

But we see that in the pure CSPIMC we are free to change $\varphi$. This is
made possible since the two ghosts for each particle are given a role 
before integrating them out. CSPIMC with different $\varphi$ will have 
different continuum limits $\tau\to 0$. This freedom allows to explore 
domains different from the one of conventional PWPIMC. And this is the 
main novelty of our CSPIMC method. Clearly, in the continuum limit $\tau\to 0$, 
the CSPIMC becomes independent of $\xi$, i.e. of the properties of the
harmonic oscillator generating the coherent states.

\begin{acknowledgments}
I would like to thank prof. Saverio Moroni for his support in the 
development of the Brownian bridge move and the consequent particles 
permutation sampling. 
\end{acknowledgments}


\section*{Author declarations}

\subsection*{Conflicts of interest}
None declared.

\subsection*{Data availability}
The data that support the findings of this study are available from the 
corresponding author upon reasonable request.

\subsection*{Funding}
None declared.

\bibliography{cspimc-new}

\end{document}